\documentclass[%
aip,
 amsmath,amssymb,
reprint,%
]{revtex4-1}

\usepackage{graphicx}
\usepackage{dcolumn}
\usepackage{bm}

\usepackage{color}

\usepackage{epstopdf}

\begin{document}


\title{Low power, fast and broadband ESR quantum control using a stripline resonator}

\author{Yung Szen Yap}
\email{yungszen@utm.my}
\affiliation{Faculty of Science and Centre for Sustainable Nanomaterials (CSNano), Universiti Teknologi Malaysia, 81310 UTM Johor Bahru, Johor, Malaysia}
\affiliation{Graduate School of Engineering Science, Osaka University, Toyonaka, Osaka 560-8531, Japan.}

\author{Makoto Negoro}
\email{negoro@ee.es.osaka-u.ac.jp}
\affiliation{Quantum Information and Quantum Biology Division, Institute for Open and Transdisciplinary Research Initiatives, Osaka University, Osaka 560-8531, Japan.}
\affiliation{JST, PRESTO, Kawaguchi, Saitama 332-0012, Japan}


\author{Mayuko Kuno}
\author{Yoshikiyo Sakamoto}
\affiliation{Graduate School of Engineering Science, Osaka University, Toyonaka, Osaka 560-8531, Japan.}

\author{Akinori Kagawa}
\affiliation{Graduate School of Engineering Science, Osaka University, Toyonaka, Osaka 560-8531, Japan.}
\affiliation{Quantum Information and Quantum Biology Division, Institute for Open and Transdisciplinary Research Initiatives, Osaka University, Osaka 560-8531, Japan.}

\affiliation{JST, PRESTO, Kawaguchi, Saitama 332-0012, Japan}

\author{Masahiro Kitagawa}
\affiliation{Graduate School of Engineering Science, Osaka University, Toyonaka, Osaka 560-8531, Japan.}
\affiliation{Quantum Information and Quantum Biology Division, Institute for Open and Transdisciplinary Research Initiatives, Osaka University, Osaka 560-8531, Japan.}

\date{\today}

\begin{abstract}
Using a home-built Ku band ESR spectrometer equipped with an arbitrary waveform generator and a stripline resonator, we implement two types of pulses that would benefit quantum computers: BB1 composite pulse and a microwave frequency comb.
Broadband type 1 (BB1) composite pulse is commonly used to combat systematic errors but previous experiments were carried out only on extremely narrow linewidth samples.
Using a sample with a linewidth of 9.35 MHz, we demonstrate that BB1 composite pulse is still effective against pulse length errors at a Rabi frequency of 38.46 MHz.  
The fast control is realized with low microwave power which is required for initialization of electron spin qubits at 0.6~T.
We also digitally design and implement a microwave frequency comb to excite multiple spin packets of a different sample.
Using this pulse, we demonstrate coherent and well resolved excitations spanning over the entire spectrum of the sample (ranging from -20 to 20 MHz).
In anticipation of scaling up to a system with large number of qubits, this approach provides an efficient technique to selectively and simultaneously control multiple qubits defined in the frequency-domain.
\end{abstract}

\keywords{EPR; ESR; AWG; composite pulse; quantum control; quantum computing; frequency comb}

\maketitle

\section{Introduction}

The recent years have seen notable progress in the quest to build universal quantum computers.\cite{arute2019quantum, heinsoo2018rapid} However, for every proposed physical systems, there are still obstacles that must be overcome in order to achieve fault-tolerant quantum computation. 
Amongst the many proposed systems, electron spin resonance (ESR) remains an interesting option for several reasons.
The electron spin is a two-level quantum system that is driven by microwave pulses in the nanosecond timescale and the electron spin system is a convenient testbed for quantum information processing experiments. In an electron-nuclear spin systems, the electron spins are suitable to be used as quantum processors
whereas the nuclear spins with its long coherence time can serve as quantum memories. \cite{mehring2006spin, hodges2008universal, morton2008solid, sessoli2018two}
With an electron gyromagnetic ratio that is approximately 660 times larger than that of proton, it is possible to prepare electron spins almost in its pure state with spin polarization exceeding 99\% at low temperatures.\cite{yap2015ku, morley2010initialization}  
Furthermore, the electron spin polarization can be transferred to initialize nuclear spins by using microwave pulses.\cite{morley2010initialization, tateishi2014room}
However, controlling electron spins with microwave pulses at such low temperatures should be carried out with as low power as possible due to the limited cooling power of the dilution refrigerator.
It is possible to achieve similar initialization levels at 1.4~K\cite{ross2001cryocoolers} and 6.8~T, but this approach operates at 190 GHz where microwave technology has yet to fully mature.

In this work, we address several issues in building a universal ESR quantum computer.  
One of the requirements is to reduce errors due to systematic inhomogeneity in a time frame that is much faster than the coherence time of the qubits (typically in the order of microseconds).\cite{divincenzo2000physical, jones2012layered} Furthermore, the physical system of the quantum computer must be scalable and the control system must be able to address specific qubits. 
Upon closer consideration, these requirements can be \emph{partly} addressed by the spectrometer itself, which is more demanding than the ones commonly used in ESR spectroscopy.

As an example, the common spectroscopy system utilizes rectangular pulses, which introduces unavoidable errors. These errors are due to inaccuracies of the pulse length, distortions to the pulse profile due to the resonator, nonlinearities of microwave components, inhomogeneity of the induced microwave magnetic field in the resonator, incorrect external magnetic field, inhomogeneous broadening of the ESR spectrum, and a long list of other sources of errors.  
While unavoidable, these errors can be minimized or corrected using specialized pulses such as composite pulses in place of the rectangular pulses.\cite{cummins2000use, cummins2003tackling, morton2005high, motion2017use} 
The composite pulses must also be much shorter than the coherence time of electron spins and would therefore require a broadband resonator to accommodate them.
Other types of pulses have also been designed for different purposes such as to overcome the limited bandwidth of the resonator,\cite{doll2014fourier, spindler2012shaped, pribitzer2016spidyan} to investigate multi-spin samples\cite{kaufmann2013dac, doll2013adiabatic, spindler2013broadband} or for quantum entanglement\cite{dolde2014high, scherer2008entangled}. 

On the other hand, in a large-scale qubit system, pulses can also be designed to selectively and simultaneously control qubits that are uniquely defined in the frequency domain. Such systems were previously demonstrated in NMR systems~\cite{vandersypen2005nmr, lloyd1993potentially} and its equivalent has been proposed for ESR systems~\cite{morita2010triple}.
These specialized pulses are sophisticated and the pulse generators or spectrometers must be able to support abrupt changes to the amplitude and phase profiles.  Such spectrometers were uncommon just around two decades ago\cite{shane1998versatile} but are recently becoming widespread\cite{tseitlin2011digital, kaufmann2013dac, yap2015ku, conway2016fpga}.

With that in mind, we performed two experiments to demonstrate fast and broad bandwidth quantum control with low microwave power. Firstly, we applied broadband type 1 (BB1) composite pulse to correct systematic errors in an experiment.  Although BB1 composite pulses have been tested before, the previous experiments were limited to samples with narrow linewidths.~\cite{morton2005high, morton2008solid, said2009robust, wu2013geometric}  
Here, we applied BB1 composite pulse on a broad linewidth sample which is a better representation of a frequency-defined quantum system.\cite{lloyd1993potentially, morita2010triple}  To faithfully apply BB1 composite pulses, we used a stripline resonator with broad bandwidth and high microwave power-to-magnetic field conversion factor (sometimes known as the resonator's efficiency).\cite{hyde1989multipurpose, yap2013strongly}
Secondly, we introduced a method to address such qubits, which is beneficial when scaling up the number of qubits of an ESR-based quantum computer.  To do this, we applied a frequency comb pulse that spanned over the entire spectrum of the sample using fast modulating shaped pulses at low power.

\section{Experimental Setup}

All of the experiments here were performed at room temperature using a home-built Ku band spectrometer described elsewhere.\cite{yap2015ku}
This spectrometer was used to generate arbitrary waveform pulses at a resolution of 0.1~ns and can be fitted to either a room temperature magnet or a dilution refrigerator for experiments at exteremely low temperatures.  

In the experiments described here, a new stripline resonator is proposed.  This resonator was somewhat similar to our previous resonator\cite{yap2013strongly} but with several notable changes (see Table~\ref{tab:resonator}). 
The new resonator was fabricated from a ceramic-filled polytetrafluoroethylene (PTFE) laminate (Rogers Duroid 6035HTC) whereas the previous resonator was fabricated from a PTFE laminate (Rogers Duroid 5880).  The new laminate was chosen for its higher thermal conductivity to take heat away from the sample.  The new laminate also featured a lower thermal expansion coefficient especially along its thickness/Z axis which prevented gaps from forming between the top and bottom layers when cooled down to millikelvin temperatures.

\begin{figure}
\centering
\begin{picture}(300,160)
\put(0,0){\includegraphics[width=\linewidth]{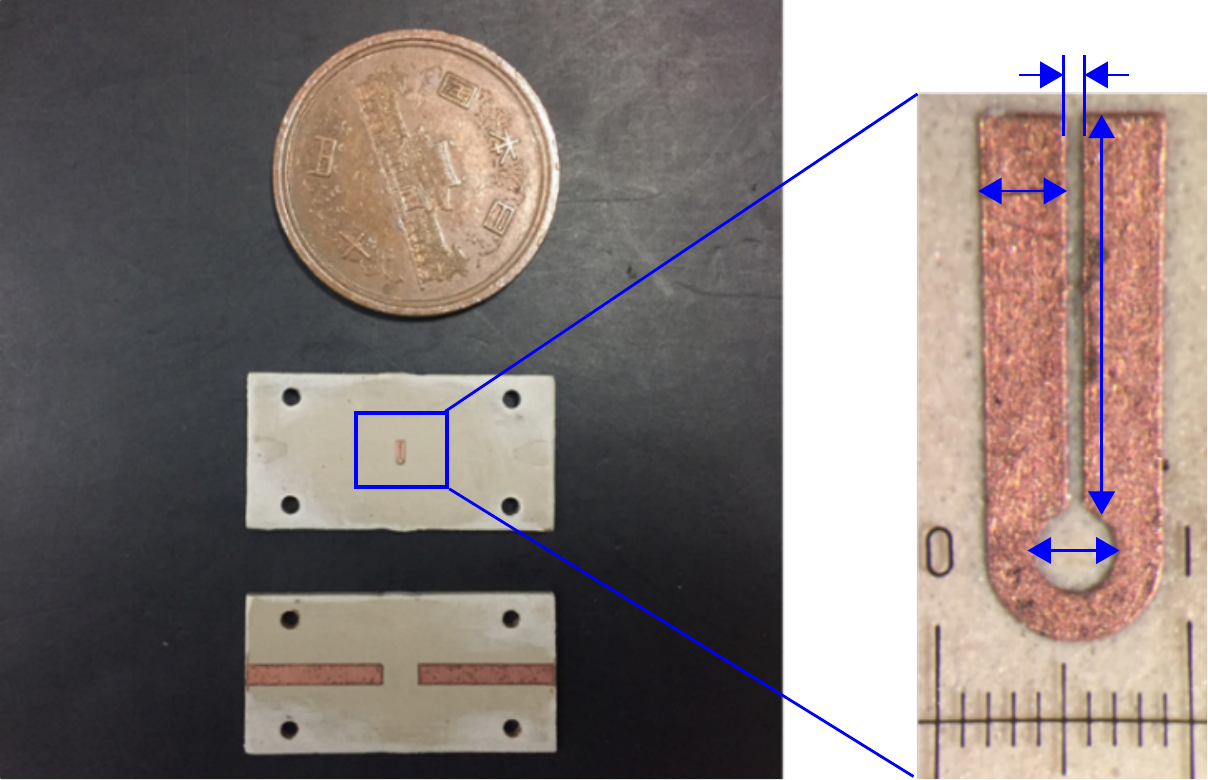}} 
\put(217,153){$g$}
\put(205,110){$w$}
\put(228,89){$h$}
\put(216,39){$d$}
\end{picture}
\caption{Photos of the fabricated resonator. The photo on the left shows the layers of stripline resonator.  The enlarged photo of resonant strip (in blue outline box) is shown on the right together with a 1 mm scale. The blue lines are used to guide readers in understanding the dimension of the resonant strip: $g$ = 0.0345 mm, $w$ = 0.35 mm, $h$ = 1.66 mm, and $d$ = 0.3 mm.  The feedline width was calculated to achieve an impedance of 50 Ohms at 17 GHz. The spacing from the feedlines to the resonant strip was 0.9 mm.}
\label{pic:resonator}
\end{figure}
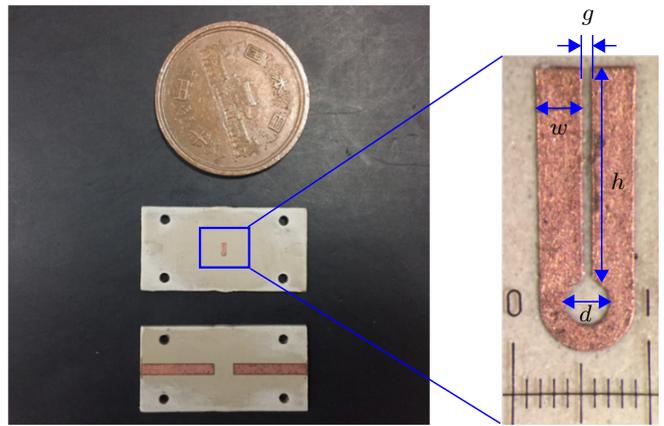

Another notable difference was the design of the resonator (Fig.~\ref{pic:resonator}), which was inspired from other sources.\cite{yap2013strongly,twig:104703}
The previous design featured a U-shape in the middle of the resonant strip with a small sample area (20 $\mu$m).\cite{yap2013strongly}  The miniaturization of the sample area and resonator led to a high resonator efficiency but also contributed to high magnetic field inhomogeneity.

The new resonator was designed for a larger sample volume with better microwave magnetic field homogeneity, but without sacrificing too much of its efficiency.  The homogeneous magnetic volume ($\pm 5\%$ from the center of the sample area) for the new resonator was estimated to be $1 \times 10^{6}$ $\mu$m$^3$, which was more than 1000 times larger than the previous design but only suffered a decrease of efficiency by a factor of about 3.5. 

Fabrication of the new resonator was done using lithography and wet-etching. A positive photoresist (OFPR 800) was applied at 500 rpm for 5 s followed by 4000~rpm for 25 s. It was prebaked at 90$^\circ\mathrm{C}$, irradiated with ultraviolet light and developed with tetramethyl ammonium hydroxide (NMD-3 2.38\%) for 25 s. It was then baked at 120$^\circ\mathrm{C}$ for 15 mins and etched with ferric chloride. The fabricated resonator has a center frequency of 17.06 GHz and a bandwidth of 255 MHz, giving a Q factor of about 66. 

\begin{table}
\centering
\caption{
Comparison between the previous\cite{yap2013strongly} and new stripline resonators.  
}

\vspace{\baselineskip}
\begin{tabular}{  m{0.5\linewidth}  c  c  }
\hline
\hline
    \\
	Design & Predecessor~\cite{yap2013strongly} & New \\ 
	\\
\hline
\\
	Diameter of sample area ($\mu$m) & 20 & 300 \\
	Thickness of Copper Layer ($\mu$m) & 20 & 17.5 \\
	\\
	Resonant frequency (GHz) & 16.8 & 17.1  \\
	Bandwidth (MHz) & 200 & 255   \\
	Q factor & 85 & 66  \\
	\\
	Experimental microwave efficiency (MHz/$\sqrt{\mathrm{W}}$) & 210 & 57.6 \\
	\\
	Thermal conductivity of substrate (Wm$^{-1}$K$^{-1}$) & 0.2~\cite{duriod5880} & 1.44~\cite{duriod6035} \\
	Coefficient of thermal expansion along Z axis (ppm/$^\circ$C) & 237 \cite{duriod5880} & 39~\cite{duriod6035} \\
\\
\hline
\hline
\end{tabular}

\label{tab:resonator}
\end{table}

Two types of samples were prepared for the experiments described here.  The first sample was (undeuterated) a,c-bisdiphe-nylene-b-phenylallyl 1:1 complex with benzene (BDPA) mixed with polystyrene, prepared according to the procedures described in Ref.~\citenum{labelle1997electron}. The second sample was deuterated BDPA, also diluted with polystyrene using the same method. The undeuterated and deuterated samples have an electron spin density of about $1.27\times10^{25}$~spins/m$^3$ and $1.2\times10^{24}$~spins/m$^3$, respectively.

\section{Experiments \& Results}

\subsection{Broadband Type 1 Composite Pulse}

In a pulsed ESR experiment, typically, a rectangular pulse is used to rotate the net magnetization vector to a desired angle, $[\theta]_\phi$. 
The notation $[\theta]_{\phi}$ represents a $\theta$ rotation pulse about an axis in the $x-y$ plane ([$\cos \phi$, $\sin \phi$, 0]). 
For example, a $[\pi/2]_{0}$ represents a 90$^\circ$ rotation operator about the $x$-axis. 
Due to imperfections and artifacts in the experiment, the desired angle cannot be achieved accurately and the rotation becomes $[\theta{}(1+\sigma)]_{\phi}$, where $\sigma$ is the fractional systematic pulse length error.\cite{cummins2003tackling}

Instead of using a single rectangular pulse, a sequence of pulses known as composite pulse can be used to achieve the desired rotation with reduced systematic error.
We employed a composite pulse known as broadband type 1 (BB1).\cite{wimperis1994broadband, cummins2003tackling, morton2005high} 
BB1 composite pulse is well-known to correct pulse length errors but is ineffective against off resonance errors. The pulse sequence for BB1 composite pulse, denoted as BB1($[\theta]_{0}$) has the following form: 
\begin{equation}
\mathrm{BB1}\left(\left[\theta\right]_0\right) =[\pi]_{\phi_1} - [2\hspace{1pt}\pi]_{\phi_2} - [\pi]_{\phi_1} - [\theta]_{0},
\label{eq:bb1}
\end{equation}

\noindent{}where the pulse sequence order is from left to right, \mbox{$\phi_1 = \cos^{-1}(-\theta/4\pi)$} and $\phi_2 = 3\hspace{1pt}\phi_1$. As an example, the BB1 pulse for $[\pi/2]_{0}$ rotation is:
\begin{equation}
\mathrm{BB1}\left(\left[\tfrac{\pi}{2}\right]_0\right) = [\pi]_{0.54\pi} - [2\hspace{1pt}\pi]_{1.62\pi} - [\pi]_{0.54\pi} - \left[\tfrac{\pi}{2}\right]_0.
\label{eq:bb1_90}
\end{equation}

Since the sample used in this experiment was an inhomogeneously broadened sample, spin echo pulse sequence was used to refocus the net magnetization and the pulse sequence has the following form:
\begin{equation}
\left[\tfrac{\pi}{2}\right]_0
- \tau
- [\pi]_0
- \tau
- (\mathrm{Echo}\hspace{1ex}\mathrm{signal}).
\label{eq:echo}
\end{equation} 

\begin{figure}
\centering
\includegraphics[width=\linewidth]{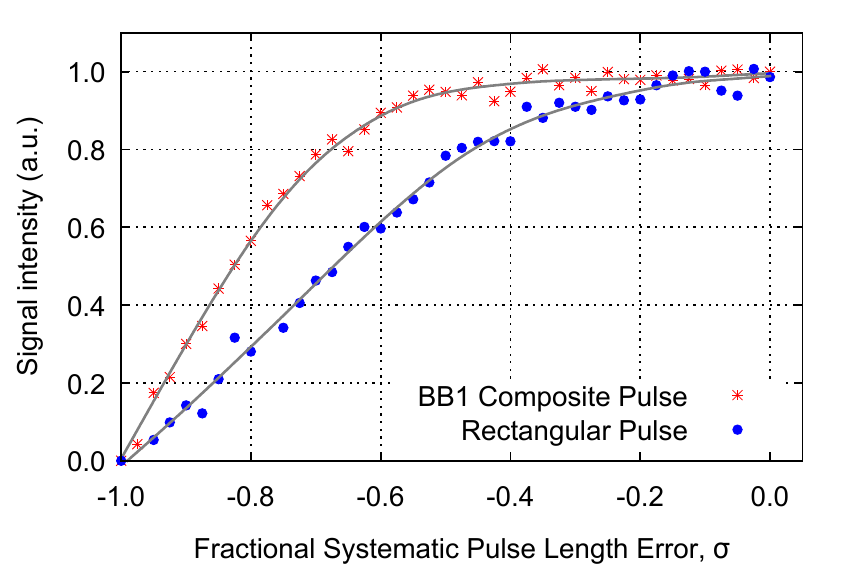}
\caption{Spin echo signal intensities obtained from normal pulses and BB1 composite pulses. The solid lines are guides for the eye. The x-axis represents the systematic pulse length error that was introduced by varying the $\pi/2$ pulse amplitude away from its optimal value.}
\label{pic:bb1_amp_results}
\end{figure}

Using deuterated BDPA diluted in PS, pulse length error was intentionally introduced into the spin echo sequence (Eq.~\ref{eq:echo}) by changing the amplitude of the $[\pi/2]_0$ pulse to: 
\begin{equation}
\left[\tfrac{\pi}{2}(1+\sigma)\right]_0
- \tau
- [\pi]_0
- \tau
- (\mathrm{Echo}\hspace{1ex}\mathrm{signal}),
\label{eq:echo_error}
\end{equation}

\noindent where $\sigma$ was caused by the deviation of the pulse amplitude away from its optimal value. This in turn deviated the net magnetization away from the intended $[\pi/2]_0$ rotation and weakened the spin echo signal intensity (see Fig.~\ref{pic:bb1_amp_results}). Next, the experiment was repeated using BB1 composite pulses:
%
\begin{multline}
[\pi]_{0.54\pi} 
- [2\hspace{1pt}\pi]_{1.62\pi} 
- [\pi]_{0.54\pi} 
- \left[\tfrac{\pi}{2}(1+\sigma)\right]_0
- \tau
\\ 
- \mathrm{BB1}([\pi]_0)
- \tau
- (\mathrm{Echo}\hspace{1ex}\mathrm{signal}).
\label{eq:bb1_amplitude}
\end{multline} 

Using Eq. \ref{eq:bb1_amplitude}, the fractional pulse length error, $\sigma$ 
was reduced and the net magnetization was rotated closer to the intended $[\pi/2]_0$ rotation angle.  This produced stronger spin echo signal intensities which only decreased to a ratio of around 0.97 even when $\sigma \approx 0.4 $ (see Fig.~\ref{pic:bb1_amp_results}). This result indicated that high fidelity rotations were achieved when BB1 composite pulses were used and the results were consistent with the simulated performance of BB1 composite pulse in Ref.~\citenum{wimperis1994broadband}.

\begin{figure}
\centering
\includegraphics[width=\linewidth]{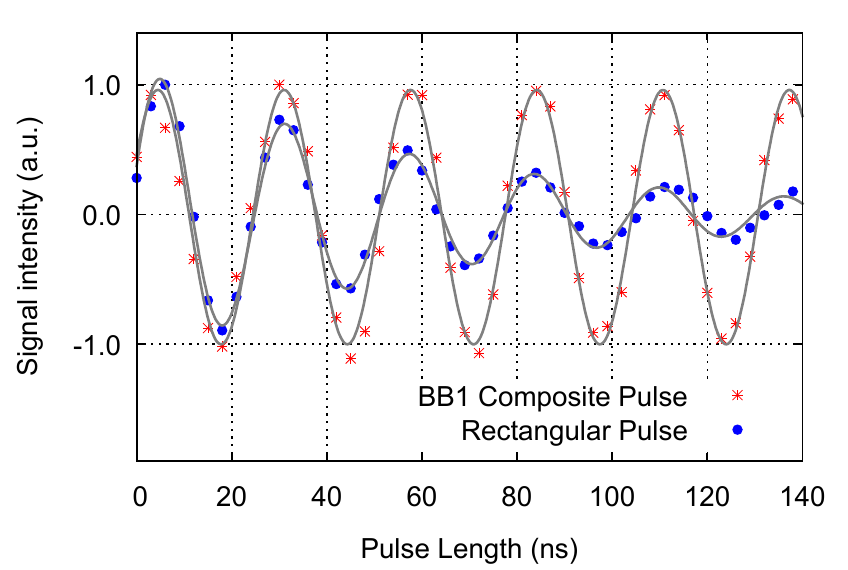}
\caption{Spin echo experiments with deuterated BDPA:PS at 0.6012 T. Spin nutation obtained using normal and BB1 pulses. In a spin nutation experiment, uncorrected systematic errors cause decoherence to build up and form a decaying sinusoidal signal which is corrected when replaced with BB1 equivalent pulses.}
\label{pic:bb1_nutation}
\end{figure}

Next, we investigated the intrinsic pulse length errors that accumulated at large rotation angles (up to $5\pi$) without introducing additional errors.
To do this, we compared the spin nutation obtained using normal and BB1 composite pulses. 
We observed that the Rabi oscillation decayed exponentially when normal pulses were used (see Fig.~\ref{pic:bb1_nutation}). 
To correct pulse length errors, the following BB1 pulse sequence was used:
\begin{multline} 
\mathrm{BB1}([\theta]_0)
- \left[\mathrm{BB1}([4\pi]_0)\right]^N
- \tau
- \mathrm{BB1}([\pi]_0)
- \tau
\\
- (\mathrm{Echo}\hspace{1ex}\mathrm{signal}),
\label{eq:bb1_4pi}
\end{multline} 
\noindent{}where $\mathrm{BB1}([\theta]_0)$ was varied from 0 to $4\pi$ and padded with $\left[\mathrm{BB1}([4\pi]_0)\right]^N$ where $N$ is the number of repetitions required to achieve rotation angles larger than $4\pi$.
By comparing the spin nutation results obtained from normal and BB1 composite pulses, we found that the exponential decay in the spin nutation was reduced significantly when BB1 pulses were used (see Fig.~\ref{pic:bb1_nutation}).  
In this experiment, since the spin-lattice relaxation time, $T_1$ was longer than the experiment time, it did not contribute to the spin nutation exponential decay. When BB1 composite pulses were used, the spin nutation decay was minimized significantly.  This confirmed that the Rabi oscillation decayed due to pulse length errors, which were corrected when BB1 composite pulses were used.

For both BB1 experiments, diluted deuterated BDPA was used. Undeuterated, pure BDPA linewidth was around 0.5--1~G (approx. 1.4--2.8~MHz) and the relaxation times were around 100~ns.\cite{goldsborough1960influence, mitchell2011electron, blank2003high} While the linewidth of undeuterated BDPA was beneficially narrow for our experiments, the short relaxation times were undesirable. 
Since BDPA narrow linewidth and its short relaxation times were due to its strong exchange interaction,\cite{goldsborough1960influence} diluting the sample weakened the exchange interaction and produced a sample with broader linewidth. 
By using a deuterated version of the diluted sample, hyperfine interaction was also reduced. As a result, the diluted deuterated sample has a spin-spin relaxation time, $T_2$ of around 200~ns and a linewidth of 9.35~MHz.
For the BB1 experiments, the pulse peak power was around 250~mW (24~dBm) and the corresponding Rabi frequency was 38.46~MHz (around 4 times higher than the linewidth of the sample).  The $\tau$ time was 300~ns and the duration of $\pi/2$, $\pi$ and $2\pi$ pulses were 6.5~ns, 13~ns and 26~ns, respectively.

\subsection{Fast, multi-frequency modulation at low power}
For molecular-spin quantum computers, well-resolved multi-frequency pulses (also known as a frequency comb) can selectively control multiple qubits defined in the frequency-space.\cite{lloyd1993potentially} In our previous work, we demonstrated a coherent, triple-frequency excitation that spanned over 8 MHz using spin echo pulses.  The pulses were consisted of a triple-frequency, shaped $\pi/2$ pulse and a rectangular $\pi$ pulse.\cite{yap2015ku} The excitation was limited by the bandwidth of the resonator, which made it insufficient for real applications. Hence, in this work, we have increased the bandwidth and number of excitations to span over the entire spectrum of the (undeuterated) diluted BDPA. 

\begin{figure}[!h]
\centering
\includegraphics[width=\linewidth]{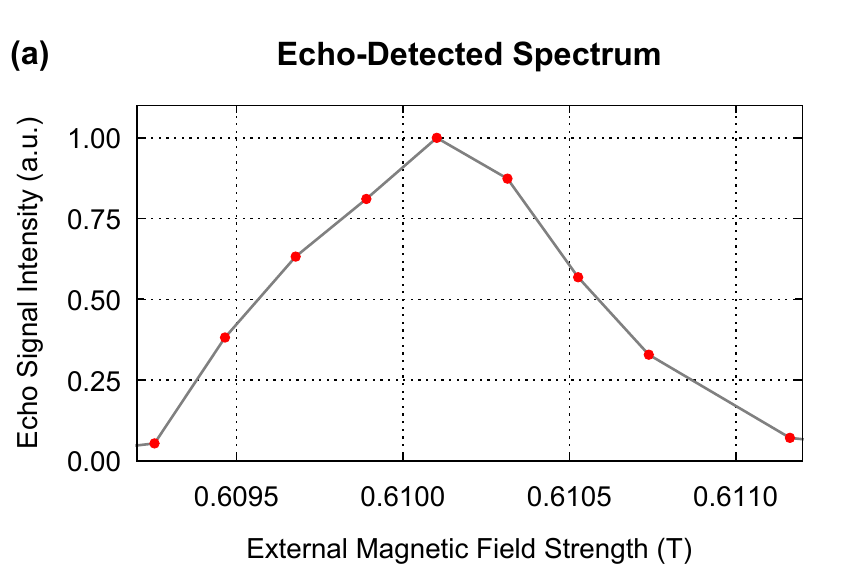}
\includegraphics[width=\linewidth]{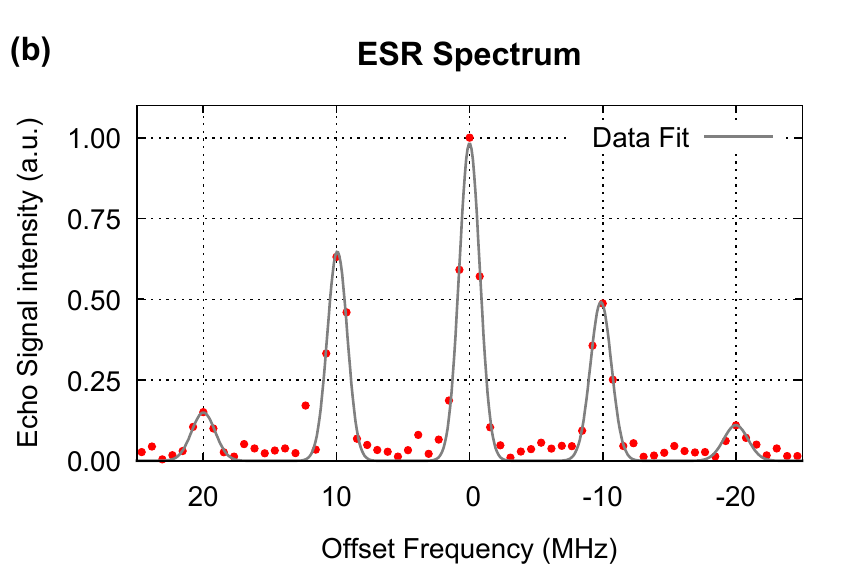}
\includegraphics[width=\linewidth]{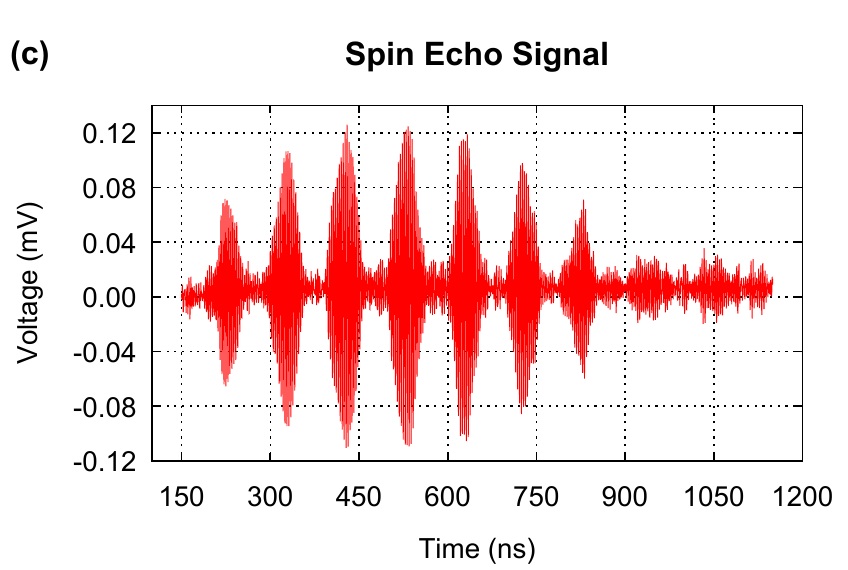}
\caption{Spin echo signals from undeuterated BDPA:PS. In (a), the echo signal intensities were obtained at different external magnetic field strength.  In (b), the spectrum was obtained using a multi-frequency shaped pulsed consisting of five offset frequencies at -20, -10, 0, 10 and 20 MHz at 0.6102 T. The pattern of the multi-frequency excitation was due to the distribution of spin packets in the sample that corresponds to (a). The corresponding spin echo signal in time domain is presented in (c), which indicates phase coherences between all frequencies.}
\label{pic:multifreq_results}
\end{figure}

Firstly, the full spectrum of the sample was obtained by sweeping the external magnetic field while an on-resonance, single-frequency Gaussian spin echo pulse was applied (see Fig.~\ref{pic:multifreq_results}~(a) for the spectrum).  
Next, the multi-frequency pulses were calculated by superimposing several single-frequency, Gaussian pulses.  This method was applied to both $\pi/2$ and $\pi$ pulses where their respective Gaussian profiles were calculated based on the duration and power of the rectangular pulses.  The respective rectangular pulse lengths for $\pi/2$ and $\pi$ pulses were 215.5 ns and 431.0 ns. The pulse peak power was 0.29 mW and the corresponding Rabi frequency was 1.16~MHz. The full-width at half maximum (FWHM) duration for a single-frequency Gaussian $\pi/2$ and $\pi$ pulses were 203.0~ns and 401.7~ns, respectively.  

Pulses consisting of five frequencies were calculated and applied at the center frequency of the spectrum (i.e. at its corresponding magnetic field).  Well-resolved excitations were obtained for all five offset frequencies: -20, -10, 0, 10 and 20 MHz, which spanned over the entire spectrum of the sample (Fig.~\ref{pic:multifreq_results}~(b)).  The interference pattern in the spin echo signal was a clear indication that the spin packets were excited coherently.  Furthermore, this was achieved with a peak power of 0.29 mW (approximately -5.4 dBm) entering the resonator (after subtracting losses from cables connecting the power amplifier to the resonator). 

\section{Summary}

We have described experiments at room temperatures using an advanced 
spectrometer capable of generating arbitrary waveform pulses and a new broad bandwidth resonator which featured a large homogeneous sample volume and without sacrificing too much efficiency. 
With this setup, we experimented with BB1 composite pulse to correct pulse length errors for a broad linewidth sample and also implemented a microwave frequency comb to excite spin packets over the entire spectrum of the sample.  Both of these experiments are important examples of arbitrary amplitude and phase modulation towards the development of a molecular spin-based quantum computer.  

For the BB1 composite pulses, two experiments were performed at a pulse power of 24 dBm (corresponding Rabi frequency was 38.46~MHz) on a sample with a linewidth of 9.35~MHz.
In the first experiment, errors were introduced by changing the $\pi/2$ pulse amplitude of the spin echo pulse sequence away from its optimal amplitude.  In the second experiment, BB1 composite pulse was applied to correct the intrinsic experimental errors (without introducing additional errors).
For both experiments, we found that the BB1 composite pulse performed well to correct pulse length errors despite the sample broad linewidth.
Previous reports of BB1 experiments were performed only on narrow linewidth ESR samples (such as N@C$_{60}$ or enriched $^{28}$Si:P) or a single spin, where off-resonance effects due to inhomogeneous broadening were negligible.~\cite{morton2005high, morton2008solid, said2009robust, wu2013geometric} 
Except for a simulation work~\cite{ishmuratov2016bulk}, as far as we know, there are no experimental reports on the effectiveness of BB1 composite pulse on a broad linewidth sample as demonstrated here.

For the multi-frequency excitation experiment, we applied a microwave frequency comb and excited five different spin packets simultaneously and coherently.
We obtained well-resolved excitations spanning over the entire spectrum of the sample ($\pm$ 20~MHz) using shaped $\pi/2$ and $\pi$ pulses. This approach is useful for controlling specific qubits without affecting the remaining qubits.  In the quest to build a universal quantum computer, controlling qubits (defined in the frequency-domain) typically involves utilizing several dedicated lines/wires for each qubit/frequency and as the number of qubits increases, the required number of control lines increases as well.~\cite{kelly2015state} This ultimately poses a scalability problem when working with a large number of qubits cooled in a dilution refrigerator. Although, the proposed approach would require additional time and effort to numerically design the multi-frequency pulses, it may provide an alternative and efficient solution to this problem.

\begin{acknowledgments}
The authors would like to thank Assoc.~Prof.~Dr.~\mbox{Chihiro}~\mbox{Yamanaka} of Osaka University for his technical assistance.
This work was supported by Japan Science and Technology Agency (JST), CREST program (JST grant number JPMJCR1672), PRESTO program (JST grant number JPMJPR1666), the Ministry of Education, Culture, Sports, Science and Technology (MEXT) Grant-in-Aid for Scientific Research on Innovative Areas 21102004, the Funding Program for World-Leading Innovating R\&D on Science and Technology (FIRST program) and JSPS KAKENHI program (Grant Number JP18H01152). Y.S.Y was also financially supported by Fundamental Research Grant Scheme (FRGS), Malaysia (FRGS/1/2018/STG02/UTM/02/15), Ministry of Education (MOE) and UTM (R.J130000.7854.5F027). The authors would like to acknowledge contributions by Dr.~Y.~Tabuchi and K.~Sowa at the early stages of the work.

\end{acknowledgments}

\bibliography{mybib}

\end{document}